\documentclass[journal,twocolumn]{IEEEtran}
\usepackage{amsfonts}
\usepackage[tbtags]{amsmath}
\usepackage{graphicx,subfigure}
\usepackage{color}      % Need the color package
\usepackage{epsfig}
\usepackage{psfig, psfrag}
\usepackage{amssymb}
\usepackage{tipa}
\usepackage{bbm}
\usepackage{algorithm2e}

\def\endproof{\hspace*{\fill}~$\blacksquare$}
\long\def\comment#1{}

\newcommand{\beq}{\begin{equation}}
\newcommand{\eeq}{\end{equation}}
\newcommand{\beqno}{\begin{equation*}}
\newcommand{\eeqno}{\end{equation*}}

\newcommand{\bes}{\begin{split}}
\newcommand{\ees}{\end{split}}
\newcommand{\bdm}{\begin{displaymath}}
\newcommand{\edm}{\end{displaymath}}

\newcommand{\goes}{\rightarrow}

\newtheorem{theorem}{Theorem}

\newtheorem{lemma}{Lemma}

\newtheorem{definition}{Definition}

\newcommand{\bd}{\begin{definition}}
\newcommand{\ed}{\end{definition}}

\newcommand{\bv}{\begin{vugraph}}
\newcommand{\ev}{\end{vugraph}}
\newcommand{\bi}{\begin{itemize}}
\newcommand{\ei}{\end{itemize}}
\newcommand{\ben}{\begin{enumerate}}
\newcommand{\een}{\end{enumerate}}

\newcommand{\bean}{\begin{eqnarray*} }
\newcommand{\eean}{\end{eqnarray*} }
\newcommand{\bea}{\begin{eqnarray} }
\newcommand{\eea}{\end{eqnarray} }
\newcommand{\beali}{\begin{align} }
\newcommand{\eeali}{\end{align} }

\newcommand{\ba}{\begin{array} }
\newcommand{\ea}{\end{array} }

\begin{document}

% paper title

\title{Communicating the sum of sources over a network}

\author{\authorblockN{Aditya Ramamoorthy\\}
\authorblockA{Department of Electrical and Computer Engineering\\
Iowa State University\\
Ames, Iowa 50011\\
Email: adityar@iastate.edu} \thanks{This research
was supported in part by NSF grant CNS-0721453.}}
% make the title area
\maketitle
\thispagestyle{empty}
\begin{abstract}
We consider a network (that is capable of network coding) with a
set of sources and terminals, where each terminal is interested in
recovering the sum of the sources. Considering directed acyclic
graphs with unit capacity edges and independent, unit-entropy sources, we show
the rate region when (a) there are two sources and $n$ terminals,
and (b) $n$ sources and two terminals. In these cases as long as
there exists at least one path from each source to each terminal
we demonstrate that there exists a valid assignment of coding
vectors to the edges such that the terminals can recover the sum
of the sources.
\end{abstract}
\vspace{-3mm}
\section{Introduction}
Network coding is a new paradigm in networking where nodes in a
network have the ability to process information before forwarding
it. This is unlike routing where nodes in a network primarily
operate in a replicate and forward manner. The problem of
multicast has been studied intensively under the paradigm of
network coding. The seminal work of Ahlswede et al. \cite{al}
showed that under network coding the multicast capacity is the
minimum of the maximum flows from the source to each individual
terminal node. The work of Li et al. \cite{lc} showed that linear
network codes were sufficient to achieve the multicast capacity.
The algebraic approach to network coding proposed by Koetter and
M\'{e}dard \cite{rm} provided simpler proofs of these results.

\par In recent years there has also been a lot of interest in the
development and usage of distributed source coding schemes due to
their applications in emerging areas such as sensor networks. Classical distributed
source coding results such as the famous Slepian-Wolf theorem
\cite{slepianwolf} usually assume a direct link between the
sources and the terminals. However in applications such as sensor
networks, typically the sources would communicate with the
terminal over a network. Thus, considering the distributed
compression jointly with the network information transfer is
important. Network coding for correlated sources was first
examined by Ho et al. \cite{tracey_ciss}. The work of Ramamoorthy
et al. \cite{adisepDSC} showed that in general separating
distributed source coding and network coding is suboptimal except
in the case of two sources and two terminals. A practical approach
to transmitting correlated sources over a network was considered
by Wu et al. \cite{WuSXK}. Reference \cite{WuSXK} also introduced
the problem of {\it Network Arithmetic} that comes up in the
design of practical systems that combine distributed source coding
and network coding.

In the network arithmetic problem, there are source nodes each of
which is observing independent sources. In addition there is a set
of terminal nodes that are only interested in the sum of these
sources i.e. unlike the multicast scenario where the terminals are
actually interested in recovering all the sources, in this case
the terminals are only interested in the sum of the sources. In
this paper we study the rate region of the network arithmetic
problem under certain special cases. In particular we restrict our
attention to directed acyclic graphs (DAGs) with unit capacity
edges and independent, unit entropy sources. Moreover, we consider the
following two cases.
\begin{itemize}
%\begin{list}{$\bullet$}{\leftmargin .01in}
\item[i)] Networks with two sources and $n$ %(for arbitrary $n$)
terminals, and \item[ii)] networks with $n$ sources %(for arbitrary
%$n$)
and two terminals.
%\end{list}
\end{itemize}
For these two cases we present the rate region for the problem.
Basically we show that as long as there exists at least one path
from each source to each terminal, there exists an assignment of
coding vectors to each edge in the network such that the terminals
can recover the sum of the sources.

This paper is organized as follows. Section
\ref{sec:net-coding-model} presents the network coding model that
we shall be assuming. Section \ref{sec:two-s-n-t} contains our
results for the case when there are two sources and $n$ terminals
and section \ref{sec:n-s-two-t} contains the results and proofs
for the case when there are $n$ sources and two terminals. In
section \ref{sec:conclusion} we outline our conclusions.
\vspace{-3mm}
\section{Network coding model}
\label{sec:net-coding-model} In our model, we represent the
network as a directed graph $G = (V,E)$. The network contains a
set of source nodes $S \subset V$ that are observing independent, discrete
unit-entropy sources and a set of terminals $T \subset V$. Our
network coding model is basically the one presented in \cite{rm}.
We assume that each edge in the network has unit capacity and can
transmit one symbol from a finite field of size $2^m$ per unit
time (we are free to choose $m$ large enough). If a given edge has
a higher capacity, it can be treated as multiple unit capacity
edges (fractional capacities can be treated by choosing $m$ large
enough). A directed edge $e$ between nodes $v_i$ and $v_j$ is
represented as $(v_i \goes v_j)$. Thus $head(e) = v_j$ and
$tail(e) = v_i$. A path between two nodes $v_i$ and $v_j$ is a
sequence of edges $\{ e_1, e_2, \dots, e_k\}$ such that $tail(e_1)
= v_i, head(e_k) = v_j$ and $head(e_i) = tail(e_{i+1}), i = 1,
\dots k-1$.

The signal on an edge $(v_i \goes v_j)$, is a linear combination
of the signals on the incoming edges on $v_i$ and the source
signal at $v_i$ (if $v_i \in S$). In this paper we assume that the
source nodes do not have any incoming edges from other nodes. If this is not the
case one can always introduce an artificial source connected to
the original source node that has no incoming edges. We shall only
be concerned with networks that are directed acyclic and can
therefore be treated as delay-free networks \cite{rm}. Let $Y_{e_i}$ (such
that $tail(e_i) = v_k$ and $head(e_i) = v_l$) denote the signal on
the $i^{th}$ edge in $E$ and let $X_j$ denote the $j^{th}$ source.
Then, we have
\begin{align*}
Y_{e_i} &= \sum_{\{e_j | head(e_j) = v_k\}} f_{j,i} Y_{e_j} \text{~if $v_k \in V \backslash S$}, \text{~and}\\
Y_{e_i} &= \sum_{\{j | X_j \text{~observed at~} v_k\}} a_{j,i} X_j
\text{~ if $v_k \in S$},
\end{align*}
where the coefficients $a_{j,i}$ and $f_{j,i}$ are from $GF(2^m)$.
Note that since the graph is directed acyclic, it is possible to
express $Y_{e_i}$ for an edge $e_i$ in terms of the sources
$X_j$'s. Suppose that there are $n$ sources $X_1, \dots, X_n$. If
$Y_{e_i} = \sum_{k=1}^n \beta_{e_i, k} X_k$ then we say that the
global coding vector of edge $e_i$ is $\boldsymbol{\beta}_{e_i} =
[ \beta_{e_i, 1} ~\cdots~ \beta_{e_i, n}]$. We shall also
occasionally use the term coding vector instead of global coding
vector in this paper. We say that a node $v_i$ (or edge $e_i$) is
downstream of another node $v_j$ (or edge $e_j$) if there exists a
path from $v_j$ (or $e_j$) to $v_i$ (or $e_i$). \vspace{-3mm}
\section{Case of two sources and $n$ terminals}
\label{sec:two-s-n-t}
In this section we state and prove the rate region for the network
arithmetic problem when there are two sources and $n$ terminals.

The basic idea of the proof is the following. We show that there
exist a certain set of nodes that can obtain both the sources
$X_1$ and $X_2$ and find a multicast code that multicasts the pair
$(X_1,X_2)$ to these nodes. We then modify the set of coding
vectors so that all the terminals can recover $X_1 + X_2$ while
ensuring that the coding vectors remain valid.

\begin{theorem}
\label{thm:2s_n_t} Consider a directed acylic graph $G = (V, E)$
with unit capacity edges, two source nodes $S_1$ and $S_2$ and $n$
terminal nodes $T_1, \dots , T_n$ such that \beqno \text{max-flow}
(S_i - T_j) \geq 1 \text{~for all~} i=1,2 \text{~and~} j=1, \dots,
n. \eeqno At each source node $S_i$, there is a unit-rate source
$X_i$. The $X_i$'s are independent. There exists an assignment of coding vectors to all edges
such that each $T_i, i = 1, \dots, n$ can recover $X_1 + X_2$.
\end{theorem}

Before embarking on the proof of this result we define a modified
graph that shall simplify our later arguments. %We perform these
%modifications in a sequences of steps.
\begin{enumerate}
\item We introduce artificial source nodes $S_1'$ and $S_2'$ such
that there exists a unit capacity edge $S_i' \goes S_i$. Similarly
we introduce artificial terminal nodes $T_i'$ and unit capacity
edges $T_i \goes T_i'$.
 Note that we are given the existence of at least one
path from $S_i \rightarrow T_j$ for all $i,j$. This in turn
implies that $\text{max-flow}(S_1' - T_j') = \text{max-flow}(S_2'
- T_j') = \text{max-flow}((S_1',S_2') - T_j') = 1$. %This is
%because the capacity of the edge $T_j \goes T_j'$ is one.
\item For each virtual terminal $T_j', j = 1, \dots, n$ there
exists a path from $S_i'$ to $T_j'$ for $i = 1,2$. Let us denote
this by $path(S_i' - T_j')$. We say that two paths intersect if
they have at least one node in common. For a given terminal
$T_j'$, in general the $path(S_1' - T_j')$ and $path(S_2' - T_j')$
could intersect in many nodes. Note that they have to intersect at
least once since the edge $T_j \goes T_j'$ is of unit capacity.
Suppose that the first intersection point is denoted $v_j$. As
demonstrated in Fig. \ref{fig:shortcut_path} it is possible to
find a new set of paths from $S_1' - T_j'$ and $S_2' - T_j'$ so
that they share the set of edges from $v_j$ to $T_j'$.

\begin{figure}[ht]
\centering

\includegraphics[width=50mm, clip=true]{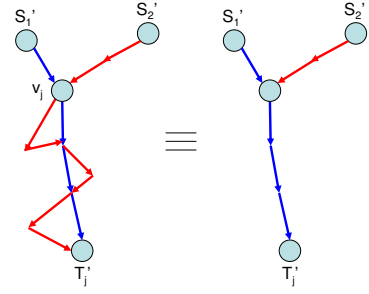} \vspace{-3mm}
\caption{\label{fig:shortcut_path} The figure on the left shows
$path(S_1' - T_j')$ (in blue) and $path(S_2' - T_j')$ (in red).
The figure on the right shows that one can find a new set of paths
from $S_1'$ and $S_2'$ to $T_j'$ such that they share edges from
$v_j$ to $T_j'$. The first intersection of the new paths is at
node $v_j$.}
\end{figure}
We assume that such paths have been found for all terminals. Thus
for each terminal $T_j'$ there exists a corresponding $v_j$ which
denotes the first vertex where the paths $S_1' - T_j'$ and $S_2' -
T_j'$ meet. Note that the $v_j$'s may not be distinct. Now,
consider the subgraph of $G$ that is defined by the union of all
these paths and suppose that we call it $G'$. In our discussion we
shall only be concerned with the graph $G'$. \item Note that $G'$
is also a directed acyclic graph. Therefore a numbering of the
nodes exists such that if there exists a path between node $v_i$
and $v_j$ then $i < j$. We now number the nodes in $G'$ in this
manner. We shall refer to the first meeting point of
$\text{path}(S_1' - T_j')$ and $\text{path}(S_2' - T_j')$ under
this new numbering as $v_{\alpha(T_j)}$.
\end{enumerate}

\begin{lemma}
In the graph $G'$ constructed as above, the following properties
hold for all $j = 1, \dots, n$.
\begin{align}
\text{max-flow}(S_1' - v_{\alpha(T_j)}) &= 1, \\
\text{max-flow}(S_2' - v_{\alpha(T_j)}) &= 1, \text{~and} \\
\text{max-flow}((S_1',S_2') - v_{\alpha(T_j)}) &= 2. \end{align}
\end{lemma}
\emph{Proof.} Obvious by the construction of the graph $G'$.
\endproof

The previous claim implies that there exists a network code so
that the pair $(X_1, X_2)$ can be multicast to each node
$v_{\alpha(T_j)}, j = 1, \dots, n$ using Theorem $8$ in \cite{rm}
. Suppose that such a network code is found and the global coding
vectors for each edge in $G'$ are found. Let these global coding
vectors be specified by the set $\boldsymbol{\beta} = \{\beta_e
~|~ e \in E'\}$.

We now present an algorithm that modifies $\boldsymbol{\beta}$ so
that each terminal $T_i', i = 1, \dots, n$ can recover $X_1 +
X_2$. This shall serve as a proof of Theorem \ref{thm:2s_n_t}.
First we sort the set $\{v_{\alpha(T_1)}, \dots,
v_{\alpha(T_n)}\}$ to obtain $\{v_{\gamma_1} , \dots ,
v_{\gamma_n}\}$ so that $\gamma_1 \leq \dots \leq \gamma_n$. Let
the terminal node corresponding to the node $v_{\gamma_i}$ be
denoted $T_{f(\gamma_i)}'$. As mentioned before it is possible
that there exist terminals $T_i$ and $T_j$ such that $\alpha(T_i)
= \alpha(T_j)$. Therefore the set of $\gamma_i$'s is not distinct.
Consequently the mapping $f(\gamma_i)$ is one to many. We do not
make this explicit to avoid the notation becoming too complex. The
steps are presented in Algorithm \ref{algo_2s_n_t}.

\incmargin{1em} \restylealgo{boxed}\linesnumbered
\begin{algorithm}[htbp]
\label{algo_2s_n_t} \SetLine Initialize $demand[i] = 0, i = 1,
\dots n$\;

\For{$k \leftarrow 1$ \KwTo $n$} {

    \If{demand[$f(\gamma_k)$] == 0}{
        \For{$e \in path(v_{\gamma_k} - T_{f(\gamma_k)}')$} {
            $\beta_e = [ 1 ~ 1]$\;
        }
        demand[$f(\gamma_k)$] = 1\;
        \For{$m \leftarrow k+1$ \KwTo $n$} {
            \If{demand[$f(\gamma_m)$] == 0} {
                \If{there exists a $path(v_{\gamma_k} - T_{f(\gamma_m)}')$} {
                    \For{$e \in path(v_{\gamma_k} - T_{f(\gamma_m)}')$} {
                        $\beta_e = [ 1 ~ 1]$\;
                    }
                    demand[$f(\gamma_m)$] = 1\;
                }
            }
        }
    }
} \caption{Algorithm for assigning coding vectors so that each
terminal can recover the sum of the two sources.}
\end{algorithm}
\decmargin{1em}  It is important to note that this algorithm
may replace the existing coding vectors assigned by the multicast
code construction on some edges. We now show that the new global
coding vector assignment is valid and is such that each terminal
receives $X_1 + X_2$.

\emph{Proof of Theorem \ref{thm:2s_n_t}.}

We claim that the assignment of coding vectors is valid at each
stage of the algorithm and by stage $1 \leq k \leq n$,
demand$[f(\gamma_k)] = 1$.
\begin{itemize}
\item {\it Base case (k=1).} Note that by the construction of $G'$
there exists a path from $v_{\gamma_1}$ to $T_{f(\gamma_1)}$. The
algorithm shall assign coding vector $[1~1]$ to those edges and
set $demand[f(\gamma_1)] = 1$. We only need to ensure that the
assignment is valid. To see the validity of the assignment note
that the graph is acyclic, therefore the coding vectors on
$path(S_1' - v_{\gamma_1})$ and $path(S_2' - v_{\gamma_1})$ do not
change. The assignments are only done on edges downstream of
$v_{\gamma_1}$ and are therefore valid.
 \item {\it Induction Step.} Assume
that the claim is true for all $j = 1, \dots, k$ and consider
stage $k+1$. If for a given $j$, the algorithm enters the for loop
on lines 4-6, we call the node $v_{\gamma_j}$ an active node.
\begin{enumerate}
\item {\it Case 1.} If there exists a path between some active
node $v_{\gamma_j}$ in the set $\{v_{\gamma_1}, \dots,
v_{\gamma_k}\}$ and $T_{f(\gamma_{k+1})}'$ then
$demand[f(\gamma_{k+1})]$ will be set to $1$ at one of the earlier
stages. By the inductive hypothesis, the assignment is valid.
\item {\it Case 2.} If $demand[f(\gamma_{k+1})]$ is still zero after
$k$ iterations of the algorithm, this implies that there does not
exist a path between an active node and $T_{f(\gamma_{k+1})}'$
i.e. there does not exist a path from an active node to any node
on $path(S_1' - T_{f(\gamma_{k+1})}')$ and $path(S_2' -
T_{f(\gamma_{k+1})}')$. Therefore the coding vectors on the edges
in $path(S_1' - T_{f(\gamma_{k+1})}') \cup path(S_2' -
T_{f(\gamma_{k+1})}')$ are unchanged at the end of iteration $k$
and are such that $v_{\gamma_{k+1}}$ receives $(X_1, X_2)$. %The
%construction of $G'$ guarantees the existence of
%$path(v_{\gamma_{k+1}} - T_{f(\gamma_{k+1})})$.
This implies that setting $\beta_e = [1 ~1]$ for $e \in
path(v_{\gamma_{k+1}} - T_{f(\gamma_{k+1})})$ will ensure that
$demand[f(\gamma_{k+1})] = 1$. This assignment is valid since the
coding vector $[1~1]$ lies in the span of the coding vector space
of $v_{\gamma_{k+1}}$. Furthermore, there does not exist a path
from $v_{\gamma_{k+1}}$ to any node on $\bigcup_{j=1}^k path(S_1'
- v_{\gamma_j}) \cup path(S_2' - v_{\gamma_j})$ since the graph is
acyclic. Therefore the assignment of coding vectors to the
previous edges remains valid. \endproof

\end{enumerate}
\end{itemize}
Note that conversely if any of the conditions in the statement of
Theorem \ref{thm:2s_n_t} is violated then there exists some
terminal that cannot obtain the value of $X_1 + X_2$. To see this
note that since the graph has unit-capacity edges the max-flow between any pair of nodes has to be an integer. Further, if for example max-flow$(S_1 - T_j) = 0$, then the
received signal at $T_j$ cannot depend on $X_1$. Since, $X_1$ and $X_2$ are independent, $X_1 + X_2$ cannot be computed at $T_j'$.

\vspace{-3mm}
\section{Case of $n$ sources and two terminals}
\label{sec:n-s-two-t} We now present the rate region for the
situation when there are $n$ sources and two terminals such that
each terminal wants to recover the sum of the sources.

To show the main result we first demonstrate that the original
network can be transformed into another network where there exists
exactly one path from each source to each terminal. This ensures
that when network coding is performed on this transformed graph
the gain on the path from a source to a terminal can be specified
by a monomial. By a simple argument it then follows that coding
vectors can be assigned so that the terminals recover the sum of
the sources.

\begin{theorem}
\label{thm:n_s_2_t} Consider a directed acylic graph $G = (V, E)$
with unit capacity edges. There are $n$ source nodes $S_1, S_2,
\dots, S_n$ and two terminal nodes $T_1$ and $T_2$ such that
\beqno \text{max-flow} (S_i - T_j) \geq 1 \text{~for all~} i = 1,
\dots, n \text{~and~}j= 1,2. \eeqno At the source nodes there are
independent unit-rate sources $X_i, i = 1, \dots, n$. There exists an
assignment of coding vectors such that each terminal can recover
the modulo-two sum of the sources $\sum_{i=1}^n X_i$.
\end{theorem}

As before we modify the graph $G$ by introducing virtual source
nodes $S_i', i = 1, \dots n$, virtual terminals $T_j', j = 1,2$
and virtual unit-capacity edges $S_i' \goes S_i, i = 1, \dots, n$
and $T_j \goes T_j', j = 1,2$. Let the set of sources be denoted
$S = \{S_1', \dots, S_n'\}$. We denote the modified graph by $G'$.
We also need the following definitions.

\begin{definition}
{\it Exactly one path condition.} Consider two nodes $v_1$ and $v_2$ such there is a path $\mathcal{P}$ between $v_1$ and $v_2$. %A path
%$\mathcal{P}$ between $v_1$ and $v_2$ is a set of edges $\{(v_1
%\goes \alpha_1), (\alpha_1 \goes \alpha_2), \dots , (\alpha_k
%\goes v_2)\} \subseteq E$.
We say that there exists exactly one
path between $v_1$ and $v_2$ if there does not exist another path
$\mathcal{P'}$ between $v_1$ and $v_2$ such that $\mathcal{P'}
\neq \mathcal{P}$.
\end{definition}
\begin{definition}
{\it Minimality.} Consider the directed acyclic graph $G'$ defined
above, with sources $S_1', \dots, S_n'$ and terminals $T_1'$ and
$T_2'$ such that
\begin{align} \label{eq:max_flow_conds} \text{max-flow}( S_i' - T_j') &= 1 ~\forall~
i = 1, \dots, n \text{~and~} j = 1,2.
\end{align}

%\begin{align} \label{eq:max_flow_conds} \text{max-flow}((\cup_{i
%\in A} S_i') - T_1') &= 1 ~\forall~ A \subseteq \{1, \dots, n\}, A
%\neq \phi
%\end{align}
The graph $G'$ is said to be minimal if the removal of any
edge from $E'$ violates one of the equalities in
(\ref{eq:max_flow_conds}).
\end{definition}

To show that Theorem \ref{thm:n_s_2_t} holds we first need an
auxiliary lemma that we state and prove.
\begin{lemma}
\label{one-path-lemma} Consider the graph $G'$ as constructed
above with sources $S_1', \dots, S_n'$ and terminals $T_1'$ and
$T_2'$. There exists a subgraph $G^*$ of $G'$ such that $G^*$ is
minimal and there exists exactly one path from $S_i'$ to $T_j'$
for $i = 1, \dots, n$ and $j = 1, 2$ in $G^*$.
\end{lemma}
\emph{Proof.} We proceed by induction on the number of sources.
\begin{itemize}
\item {\it Base case $n=1$.} In this case there is only one source
$S_1'$ and both the terminals need to recover $X_1$. Note that we
are given the existence of $path(S_1'-T_1')$ and $path(S_1'-T_2')$
in $G'$. In general these paths can intersect at multiple nodes
which may imply that there exist multiple paths (for example) from
$S_1'$ to $T_1'$. Now, from $path(S_1'-T_1')$ and
$path(S_1'-T_2')$ we can find the last node where these two paths
meet. Let this last node be denoted $u_1$. Then as shown in Fig.
\ref{fig:shortcut_path_2} we can find a new set of paths from
$S_1'$ to $T_1'$ and $S_1'$ to $T_2'$ that overlap from $S_1'$ to
$u_1$ and have no overlap thereafter. Choose $G^*$ to be the union
of these new set of paths. It is easy to see that in $G^*$ there
is exactly one path from $S_1'$ to $T_1'$ and exactly one path
from $S_1'$ to $T_2'$. Moreover removing any edge from $G^*$ would
cause at least one path to not exist. \vspace{-2mm}
\begin{figure}[ht]
\centering
\includegraphics[width=50mm, clip=true]{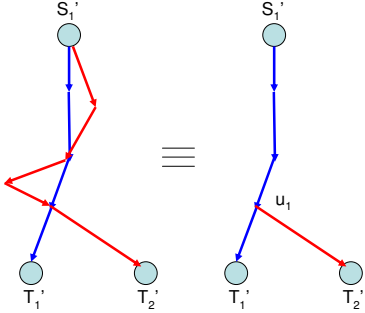}
\caption{\label{fig:shortcut_path_2} The figure on the left shows
$path(S_1' - T_1')$ (in blue) and $path(S_1' - T_2')$ (in red).
The figure on the right shows that one can find a new set of paths
from $S_1'$ to $T_1'$ and $T_2'$ such that they share edges from
$S_1'$ to $u_1$ and have no intersection thereafter.}
\end{figure}
\vspace{-3mm} \item {\it Induction Step.} We now assume the
induction hypothesis for $n-1$ sources. i.e. there exists a
minimal subgraph $G^*_{n-1}$ of $G'$ such that there is exactly
one path from $S_i'$ to $T_j'$ for $i = 1, \dots , n-1$ and $j =
1,2$. Using this hypothesis we shall show the result in the case
when there are $n$ sources.

As a first step color the edges in the subgraph $G^*_{n-1}$, blue
(the remaining edges in $G'$ have no color). The conditions on
$G'$ guarantee the existence of $path(S_n' - T_1')$ and $path(S_n'
- T_2')$. Note that these paths may intersect at many nodes. We
preprocess them in the following manner. Find the last node not in
$G^*_{n-1}$ belonging to both $path(S_n' - T_1')$ and $path(S_n' -
T_2')$. Suppose that this node is denoted $v_{r}$. Find a new set
of paths such that they share edges from $S_n'$ to $v_r$ and call
these new paths $path(S_n' - T_1')$ and $path(S_n' - T_2')$. Color
all edges on $path(S_n' - T_1')$ and $path(S_n' - T_2')$ red. This
would imply that some edges have a pair of colors. Now, consider
the subgraph induced by the union of the blue and red subgraphs
that we denote $G_{br}$.

Find the first node at which $path(S_n' - T_1')$ intersects the
blue subgraph and call that node $u_1$. Similarly find the first
node at which $path(S_n' - T_2')$ intersects the blue subgraph and
call that node $u_2$.

Observe that in $G^*_{n-1}$ there has to exist a $path(S_i' -
T_j')$ for some $i = 1, \dots, n-1$ and $j=1,2$ that passes
through $u_1$. To see this assume otherwise. This implies that
$u_1$ does not lie on any path connecting one of the sources to
one of the terminals. Therefore the incoming and the outgoing
edges of $u_1$ can be removed without violating the max-flow
conditions in (\ref{eq:max_flow_conds}). This contradicts the
minimality of $G^*_{n-1}$. Therefore we are guaranteed that there
exists at least one source such that there exists an exclusively
blue path from it to $u_1$ in $G^*_{n-1}$. A similar statement
holds for the node $u_2$. We now establish the statement of the
lemma when there are $n$ sources.

\begin{itemize}
\item {\it Case 1.} In $G_{br}$ there exists a path from $u_1$ to
$T_2'$ such that all edges on this path have a blue component.\\

First, we remove the color red from all edges on $path(S_n' -
T_2') \backslash path(S_n' - T_1')$. Next, form a subset of the
sources denoted $S^{(u_1)}$ in the following manner. For each
source $S_i', i = 1, \dots, n$ do the following.
\begin{itemize}
\item[i)] If there exists a path (with edges of color red or blue)
from $S_i'$ to $u_1$, add it to set $S^{(u_1)}$ \footnote{A path
from $S_i'$ to $u_1$ cannot have a (red,blue) edge since $u_1$ is
the first node where a red path intersects the blue subgraph}.
\end{itemize}
Let $G^{(u_1)}$ denote the subgraph induced by $\bigcup_{S_i' \in
S^{(u_1)}} path(S_i' - u_1)$.

\par Consider the graph obtained by removing the subgraph
$G^{(u_1)}$ from $G_{br}$. We denote this graph $G_{br}^-$. We
claim that the max-flow conditions in (\ref{eq:max_flow_conds})
continue to hold over $G_{br}^-$ for the set of sources $S
\backslash S^{(u_1)}$. Furthermore there still exist $path(u_1 -
T_1')$ and $path(u_1 - T_2')$ in $G_{br}^-$.

To see this note that the max-flow conditions for a source $S_i'
\in S \backslash S^{(u_1)}$ can be violated only if an edge $e$
belonging to a path from $S_i'$ to $T_j', j =1,2$ is removed. This
happens only if there exists a path from $e$ to $u_1$ which
contradicts the fact that $S_i' \in S \backslash S^{(u_1)}$. Next,
there still exist paths from $u_1$ to the terminals since the
edges on these paths are downstream of $u_1$. If any of these was
removed by the procedure, this would contradict the acyclicity of
the graph.

Note that the subgraph $G^{(u_1)}$ contains a set of sources
$S^{(u_1)}$ and a single node $u_1$ such that there exists exactly
one path from each source in $S^{(u_1)}$ to $u_1$. This has to be
true for the sources in $S^{(u_1)} \backslash \{S_n\}$ otherwise
the minimality of $G^*_{n-1}$ would be contradicted and is true
for $S_n$ by construction.

Next, introduce an artificial source $S_{a}$ and an edge $S_a
\goes u_1$ in $G_{br}^-$. Note that $|S \backslash S^{(u_1)}| \leq
n-2$, which means that the total number of sources in $G_{br}^-$
(including $S_a$) is at most $n-1$. Therefore the induction
hypothesis can be applied on $G_{br}^-$ i.e. there exists a
subgraph of $G_{br}^-$ such that there exists exactly one path
from $(S \backslash S^{(u_1)}) \cup \{S_a\}$ to each terminal.
Suppose that we find this subgraph. Now remove $S_a$ and the edge
$S_a \rightarrow u_1$ from this subgraph and augment it with the
subgraph $G^{(u_1)}$ found earlier. We claim that the resulting
graph has the property that there exists exactly one path from
each source to each terminal.

To see this note that there exists only one path from a source
$S_i' \in S \backslash S^{(u_1)}$ to $T_j', j = 1,2$. This is
because even after the introduction of $G^{(u_1)}$ there does not
exist a path from $S_i'$ to $u_1$ in this graph. Therefore the
introduction of $G^{(u_1)}$ cannot introduce additional paths
between $S_i' \in S \backslash S^{(u_1)}$ and the terminals. Next
we argue for a source $S_i' \in S^{(u_1)}$. Note that there exists
exactly one path from $u_1$ to both the terminals so the condition
can be violated only if there exist multiple paths from $S_i' \in
S^{(u_1)}$ to $u_1$, but the construction of $G^{(u_1)}$ rules
this out.

\item {\it Case 2.} In $G_{br}$ there exists a path from $u_2$ to
$T_1'$ such that all edges on this path have a blue component.\\

This case can be handled in exactly the same manner as in case 1
by removing the color red from all edges on $path(S_n' - T_1')
\backslash path(S_n' - T_2')$ and applying similar arguments for
$u_2$.

\item {\it Case 3.} In $G_{br}$ there (a) does not exist a path
with blue edges from $u_1$ to $T_2'$, and (b) does not exist a
path with blue edges from $u_2$ to $T_1'$.\\

As shown previously $u_1$ lies on some path from $S_i'$ to $T_j'$
for some $i$ and $j$ in $G^*_{n-1}$. In the current case there
does not exist a blue path from $u_1$ to $T_2'$. Therefore there
has to exist a blue path from $u_1$ to $T_1'$ in $G^*_{n-1}$. A
similar argument shows that there has to exist a blue path from
$u_2$ to $T_2'$ in $G^*_{n-1}$.

Note that the exclusively red paths from $S_n'$ to $u_1$ and $u_2$
are such that they overlap until their last intersection point.
Now, choose the desired subgraph to be the union of $G^*_{n-1}$
and the red paths, $path(S_n' - u_1)$ and $path(S_n' - u_2)$ i.e.
$G^*_n = G^*_{n-1} \cup path(S_n' - u_1) \cup path(S_n' - u_2)$.
By the induction hypothesis there exists exactly one path between
$S_i', i = 1, \dots, n-1$ and $T_j', j = 1,2$. This continues to
be true in $G^*_n$, since the red edges cannot be reached from the
blue edges. To see that there is exactly one path from $S_n'$ to
$T_1'$, assume otherwise and observe that there is exactly one
path from $S_n'$ to $u_1$ by the construction of the red paths.
Thus the only way there can be multiple paths from $S_n'$ to
$T_1'$ is if there are multiple paths from $u_1$ to $T_1'$, but
this would contradict the induction hypothesis since this would
imply that there exists some $S_i', i = 1, \dots, n-1$ that has
multiple paths to $T_1'$. A similar argument shows that there
exists exactly one path from $S_n'$ to $T_2'$.
\endproof
\end{itemize}
\end{itemize}

%We now present the proof of Theorem \ref{thm:n_s_2_t}. \\
\emph{Proof of Theorem \ref{thm:n_s_2_t}.} From Lemma
\ref{one-path-lemma} we know that it is possible to find a
subgraph $G^*$ of $G$ such that there exists exactly one path from
$S_i'$ to $T_j'$ for all $i=1, \dots, n$ and $j=1,2$. Suppose that
we find $G^*$. We will show that each terminal can recover
$\sum_{i=1}X_i$ by assigning appropriate local encoding
responsibilities for every node. Consider a node $v \in G^*$ and
let $\Gamma^o(v)$ and $\Gamma^i(v)$ represent the set of outgoing
edges from $v$ and incoming edges into $v$ respectively. Let $Y_e$
represent the symbol transmitted on edge $e$. Each node operates
in the following manner.
\begin{align}
Y_e &= \sum_{e' \in \Gamma^i(v)} \alpha \times Y_{e'} \text{~for
$e \in \Gamma^o(v)$}
\end{align}
i.e. each node scales the symbol on each input edge by $\alpha$
(note that $\alpha$ is the same for every node) and the forwards
the sum of the scaled inputs on all output edges. We shall see
that the setting $\alpha = 1$ will ensure that each terminal
recovers $\sum_{i=1}^n X_i$. To see this we examine the transfer
matrix from the inputs $[X_1 ~ \dots ~ X_n]_{1 \times n}$ to the
output $Z_{T_j \goes T_j'}$ denoted $M_j$ which is of dimension $n
\times 1$ i.e. $Z_{T_j \goes T_j'} = [X_1 ~ \dots ~ X_n] M_j$.
Note that the $i^{th}$ entry of $M_j$ corresponds to the sum of
the gains from all possible paths from $S_i'$ to $T_j'$. The
construction of $G^*$ ensures that there is exactly one such path.
Therefore the $i^{th}$ entry of $M_j$ will be a non-zero monomial
in $\alpha$ for all $i = 1, \dots, n$. Now setting $\alpha = 1$
will ensure that all the monomials evaluate to $1$ i.e. $M_j = [1
~ \cdots ~ 1]$, which implies that $Z_{T_j \goes T_j'} =
\sum_{i=1}^n X_i$. \endproof

As in the previous section it is clear that if any of the
conditions in the statement of Theorem \ref{thm:n_s_2_t} are
violated then either terminal $T_1$ or $T_2$ will be unable to
find $\sum_{i=1}^n X_i$. For example if max-flow$(X_j - T_1) = 0$
then the received signal at $T_1$ cannot depend on $X_j$. Thus,
$T_1$ cannot compute any function that depends on $X_j$.
\vspace{-2mm}
\section{Conclusion}
\label{sec:conclusion} We considered the problem of finding the
rate region for the problem of communicating the modulo-2 sum of a
set of independent unit rate sources to a set of terminals in the case when
the underlying network can be modeled as a directed acyclic graph
with unit capacity edges. The rate region has been presented for
the cases when there are (a) two sources and $n$ terminals, and
(b) $n$ sources and two terminals. 
Rate regions for arbitrary number of sources and terminals over
general network topologies possibly containing cycles are
currently under investigation. \vspace{-2mm}
\bibliographystyle{IEEEtran}
\bibliography{tip}

% Generated by IEEEtran.bst, version: 1.12 (2007/01/11)
\begin{thebibliography}{1}
\providecommand{\url}[1]{#1}
\csname url@samestyle\endcsname
\providecommand{\newblock}{\relax}
\providecommand{\bibinfo}[2]{#2}
\providecommand{\BIBentrySTDinterwordspacing}{\spaceskip=0pt\relax}
\providecommand{\BIBentryALTinterwordstretchfactor}{4}
\providecommand{\BIBentryALTinterwordspacing}{\spaceskip=\fontdimen2\font plus
\BIBentryALTinterwordstretchfactor\fontdimen3\font minus
  \fontdimen4\font\relax}
\providecommand{\BIBforeignlanguage}[2]{{%
\expandafter\ifx\csname l@#1\endcsname\relax
\typeout{** WARNING: IEEEtran.bst: No hyphenation pattern has been}%
\typeout{** loaded for the language `#1'. Using the pattern for}%
\typeout{** the default language instead.}%
\else
\language=\csname l@#1\endcsname
\fi
#2}}
\providecommand{\BIBdecl}{\relax}
\BIBdecl

\bibitem{al}
R.~Ahlswede, N.~Cai, S.-Y. Li, and R.~W. Yeung, ``{Network Information Flow},''
  \emph{IEEE Trans. on Info. Th.}, vol. 46, no. 4, pp. 1204--1216, 2000.

\bibitem{lc}
S.-Y. Li, R.~W. Yeung, and N.~Cai, ``{Linear Network Coding},'' \emph{IEEE
  Trans. on Info. Th.}, vol. 49, no. 2, pp. 371--381, 2003.

\bibitem{rm}
R.~Koetter and M.~M\'{e}dard, ``{Beyond Routing: An Algebraic Approach to
  Network Coding},'' in \emph{IEEE Infocom}, 2002.

\bibitem{slepianwolf}
D.~Slepian and J.~Wolf, ``{Noiseless coding of correlated information
  sources},'' \emph{IEEE Trans. on Info. Th.}, vol.~19, pp. 471--480, Jul.
  1973.

\bibitem{tracey_ciss}
T.~Ho, M.~M\'{e}dard, M.~Effros, and R.~Koetter, ``{Network Coding for
  Correlated Sources},'' in \emph{CISS}, 2004.

\bibitem{adisepDSC}
A.~Ramamoorthy, K.~Jain, P.~A. Chou, and M.~Effros, ``{Separating Distributed
  Source Coding from Network Coding},'' \emph{IEEE Trans. on Info. Th.},
  vol.~52, pp. 2785--2795, June 2006.

\bibitem{WuSXK}
Y.~Wu, V.~Stankovi\'c, Z.~Xiong, and S.~Y. Kung, ``{On practical design for
  joint distributed source coding and network coding},'' in \emph{Proceedings
  of the First Workshop on Network Coding, Theory and Applications, Riva del
  Garda, Italy}, 2005.

\end{thebibliography}
\end{document}